\documentclass[12pt]{iopart}
\usepackage{iopams}
\usepackage[dvips]{graphicx}

\begin{document}

\title[Stochastic coupling in two modes systems ...]{Stochastic coupling in two modes systems: the weak-strong coupling transition}

%\shorttitle{Title} %Insert here a short version of the title if it exceeds 70 characters

%\shortauthor{J. G. G. de Oliveira Jr \etal}

\author{Dagoberto S. Freitas$^{1,2}$\footnote{Author to whom any
correspondence shoud be addressed.\\ Permanent addres: Departamento
de F\'\i sica, Universidade Estadual de Feira de Santana, 44036-900,
Feira de Santana, BA, Brazil.} and M. C. Nemes$^1$}

\address{$ˆ1$ Departamento de F\'\i sica,
Instituto de Ci\^encias Exatas, Universidade Federal de Minas
Gerais, 30123-970, Belo Horizonte, MG, Brazil}
\address{$^2$ Departamento de F\'\i
sica, Universidade Estadual de Feira de Santana, 44036-900, Feira de
Santana, BA, Brazil}

\ead{dfreitas@uefs.br}

%\date{\today}

\begin{abstract}
We investigate the weak-strong coupling transition of two linearly
coupled systems under the influence of a phase fluctuating coupling.
In the weak coupling regime the exponential decay of quantum
properties is well known. A different scenario occurs in the strong
coupling regime, the inhibition of the dynamics which tends to
``freeze" as the ration between coupling strength and average phase
fluctuation time increase. Exciton-polariton oscillations and the
self-trapping phenomenon in Bose-Einstein Condensate qualitatively
illustrate the weak and strong regimes respectively.
\end{abstract}


\maketitle

\section{Introduction}
Decoherence effects are now believed to be the essential ingredient
which destroys most of the counterintuitive aspects of quantum
mechanics. Such effects are at the same time an academic tool to the
understanding of the classical limit of quantum mechanics as well as
an important ingredient in the area of quantum computation. The
dynamics of quantum open systems has been therefore extensively
studied \cite{legget}. Of particular importance in this context is
the Born-Markov approximation which leads to master equations of
various kinds \cite{angel10}, whose validity is limited by the weak
coupling approximation. The strong coupling regime however has been
less explored.

It is the purpose of the present contribution to shed some light
onto the weak-strong coupling transition in the context of two
linearly interacting systems under the influence of a phase
fluctuating coupling. The model in spite of its schematic character
has been shown in several instances and different areas to reflect
and adequately describe experimental results. Examples are the
description of exciton-polariton damped oscillations \cite{jacob95,
pau95}; predictions for the behavior of oscillations of two coupled
modes in the context of microwave cavities \cite{haroche08,
rossi09}; the self-trapping phenomenon in the tunneling process of a
Bose-Einstein condensate (BEC) \cite{michael05, milburn97, smerzi97,
ruostekoski98, meier01, kalosaka02, kalosaka03, salgueiro07}.

We will show that in the weak coupling regime the usual master
equation results are recovered and the usual phenomenological
damping constant is derived as a function of the model parameters.
The strong coupling limit however leads to a completely different
scenario, the ``freezing" of the dynamics. We will illustrate these
effects in the context of exciton-polariton oscillations and the
self-trapping of a BEC in a devised laser potential.

\section{The Model}
In this section we give a detailed derivation of the stochastic time
evolution of the following system
\begin{eqnarray}
H  = \hbar\omega_{a}a^{\dag}a + \hbar\omega_{b}b^{\dag}b +
\hbar\big[g(t)a^{\dag}b + g^{*}(t)b^{\dag}a\big],\label{eqH}
\end{eqnarray}
where $a^{\dag} (a)$ and $b^{\dag} (b)$ are creation (annihilation)
operators. $\omega_{a}$ and $\omega_{b}$ are the frequencies of the
a-mode and b-mode respectively. Since that the operators $a$ and $b$
were considered as bosons they will obey the commutation relation
for bosons $[a, a^{\dag}] = 1$ and $[b, b^{\dag}] = 1$. In the third
term, $g(t)$ stands for the interaction strength between the modes
and here is assumed to be time dependent in the sence that $|g(t)|$
is constant but its phase is a stochastic variable. The Hamiltonian
(\ref{eqH}) can be written in matrix form
\begin{eqnarray}
H & = & \hbar\left(\begin{array}{cc} a^{\dag} & b^{\dag}\end{array}
\right)\left(
          \begin{array}{cc}
            \omega_{a} & g(t) \\
            g^{*}(t) & \omega_{b} \\
          \end{array}
        \right)\left(\begin{array}{cc} a \\ b\end{array}
\right). \label{Hmtx}
\end{eqnarray}
The time evolution operator for the system is given by
\begin{eqnarray}
U(t) & = & e^{-iH_{int}t/\hbar}\nonumber\\
& = & \left(
          \begin{array}{cc}
            \cos\big(|g|t\big) & \frac{g(t)}{|g|}\sin\big(|g|t\big) \\
            -\frac{g^{*}(t)}{|g|}\sin\big(|g|t\big) & \cos\big(|g|t\big) \\
          \end{array}
        \right),
\label{Umtx}
\end{eqnarray}
where $H_{int}$ is the interaction part of the Hamiltonian $H$.

\begin{figure}[h]
\centering
\includegraphics[scale=0.35]{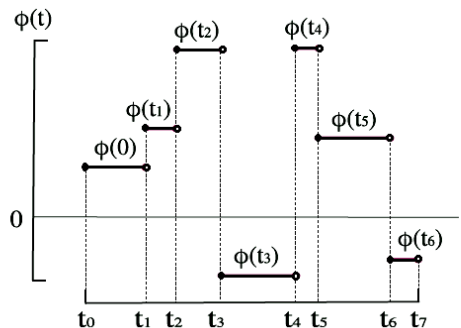}
\caption{Schematic representation of the phase distribution of
coupling between the modes.} \label{fig1}
\end{figure}

We will specify the noise by defining a stochastic process for
$\phi(t)$. In our model we assume that
\begin{equation}
g(t)=g_{0}\exp\big[i\phi(t)\big], \label{gt}
\end{equation}
where $g_{0}$ is the non-stochastic amplitude while the phase
$\phi(t)$ is treated as a stochastic variable. Here, we will
consider random phase telegraph noise where $\phi(t)$ itself
fluctuates in the manner of jumps. In particular, the phase
fluctuations were describe by a Wierner-Levy (phase diffusion)
process and the amplitude fluctuations by a colored gaussian noise.
An alternative model which represents noise by means of discrete
jump processes was first introduced into quantum optics by Burshtein
and Oseledchik \cite{bursh66}. A simple example of such a jump
process is the two-state random telegraph. These models are very
convenient and elegant to study the noise of the electromagnetic
atom-field interaction in a non-perturbative manner. The random
telegraph models, whether associated with phase, frequency or
amplitude fluctuations lead to an equation for average responses in
exact algebraic form. The model of random telegraph (jump-type)
noise is physically very sound to describe the noise arising from
electromagnetic field fluctuation or from collisions of various
kinds or from other external sources. Indeed, that model including
the effects of stochastic phase and/or in amplitude has been
explicitly solved for the case of the James-Cumming model (JCM) by
A. Joshi \cite{joshi95}. The fluctuations are modeled by the random
telegraph process and an equation for the density operator averaged
over the fluctuations is obtained. The solution of these equations
was used to study the decoherence effects in the dynamics of the
system. A. Joshi's work was treated in a pedagogical form by E. A.
Ospina \cite{ospina09}. We assume further that the change in
$\phi(t)$ occurs instantaneously jump wise and the jumps are
separated by mean time intervals of the order $\tau_{0}$ in which
$\phi(t) =$ constant, as shown in Fig. (\ref{fig1}).

There are two stochastic variables: the time interval $\tau$ between
one and the next jump and the value of the phase  constant $\phi$ in
each of these intervals. The variable $\tau = t_{i}-t_{i-1}$ follow
the probability distribution
\begin{equation}
dQ(\tau) = \frac{1}{\tau_{0}}e^{-\tau/\tau_{0}}d\tau.\label{eqdQ}
\end{equation}
with $t_{0}=0$. The above distribution specifies the probability of
duration of each such jump interval and has intervals mean duration
$\tau_{0}$. We consider only the case in which the phases $\phi(t)$
are uncorrelated. The probability distribution to phase is given by
\begin{equation}
dq(\phi) = \frac{d\phi}{2\pi},
\end{equation}
with mean value $\langle\phi\rangle = \pi$. So, at any instant, the
probability of finding a given $\phi$ remains the same and equation
to $dQ(t)$, and there is no limitation on the form of this
distribution. In other words, $\phi(t)$ is undergoing random
continuous change of Markov type.

The dynamics of the system is given by the unitary transformation
$U(\phi,t,t^{'})$ such that
\begin{equation}
\rho(t;\phi)=U(\phi,t,t^{'})\rho(t^{'})U^{-1}(\phi,t,t^{'}).\label{eqrho}
\end{equation}
At the end of each (\emph{i}th) interval we find the density matrix
$\rho(t)$ which is the initial condition for the next matrix, so, if
in the interval $(0,t)$ there are $k$ jumps in $\phi$, then
\begin{eqnarray}
\rho(t;t_{1},\cdots, & t_{k} &,\phi_{0},\cdots,\phi_{k}) =
U(\phi_{k};t,t_{k})U(\phi_{k-1};t_{k},t_{k-1})\cdots\nonumber\\
& \times &  U(\phi_{1};t_{2},t_{1})U(\phi_{0};t_{1},0)
\rho(0)U^{-1}(\phi_{0};t_{1},0)U^{-1}(\phi_{1};t_{2},t_{1})\cdots\nonumber\\
& \times &
U^{-1}(\phi_{k-1};t_{k},t_{k-1})U^{-1}(\phi_{k};t,t_{k}).\label{eqrhot1}
\end{eqnarray}
The above expression is of multiplicative nature and hence it is
quite easy to average over. The probability (in the interval
$(0,t)$) that $k$ changes of $\phi$ have actually occurred at
successive instants $t_{1},t_{2},\cdots,t_{k}$ and that a certain
sequence of $\phi_{1},\phi_{2},\cdots,\phi_{k}$ (where $\phi_{i} =
\phi(t_{i})$) was realized between them is obviously equal to
\begin{eqnarray}
dP(t_{1},t_{2},\cdots,t_{k};\phi_{1},\phi_{2},\cdots,\phi_{k},t) =
\frac{1}{\tau^{k}_{0}}e^{-t/\tau_{0}}\Big(
\prod_{i=1}^{k}dt_{i}\Big)\Big(
\prod_{i=1}^{k}dq(\phi_{i})\Big).\label{proba}
\end{eqnarray}
The average density operator can thus be written as
\begin{eqnarray}
\bar{\rho}(t) = \sum_{k=0}^{\infty}\int\int
dP(t_{1},t_{2},\cdots,t_{k};\phi_{1},\phi_{2},\cdots,\phi_{k},t)\rho(t;t_{1},\cdots,
& t_{k} &,\phi_{0},\cdots,\phi_{k}).\nonumber\\
\label{eqrho_comp}
\end{eqnarray}
Rewrite (\ref{eqrho_comp}) with use of (\ref{proba}) we get
\begin{eqnarray}
\bar{\rho}(t)e^{t/\tau_{0}} & = &
\sum_{k=0}^{\infty}\frac{1}{\tau^{k}_{0}}\int_{0}^{t}dt_{k}
\int_{0}^{t_{k}}dt_{k-1}\cdots\int_{0}^{t_{2}}dt_{1}\int
dq(\phi_{k})\int dq(\phi_{k-1})\cdots\nonumber\\
& \times &\int dq(\phi_{0})\rho(t;t_{1},\cdots,t_{k}
,\phi_{0},\cdots,\phi_{k}).\label{eqrhot2}
\end{eqnarray}
Note that the term with $k=0$ (when $\phi$ does not change at all in
the interval $(0,t)$) will not contain integrals with respect to
time and thus is given by
\begin{equation}
\int dq(\phi_{0})\rho(t,\phi_{0}).
\end{equation}
Now using the recurrence relation (\ref{eqrho}), we can multiply
both sides of equation (\ref{eqrhot2}) from the left (right) by
$U^{-1}(\phi;\tau,t)$ ($U(\phi;\tau,t)$) respectively and also by
$dq(\phi)dt/\tau_{0}$, then integrate with respect to time from $0$
to $\tau$ and eliminate the entire series using equation
(\ref{eqrhot2}). After some simplifications it is easy to show that
\begin{eqnarray}
\bar{\rho}(\tau)e^{\tau/\tau_{0}} & = &
\int dq(\phi_{0})U(\phi_{0};\tau,0)\rho(0)U^{-1}(\phi_{0};\tau,0)\nonumber\\
& + & \frac{1}{\tau_{0}}\int_{0}^{\tau}dt e^{t/\tau_{0}}\int
dq(\phi)
U(\phi;\tau,t)\bar{\rho}(t)U^{-1}(\phi;\tau,t).\label{eqrhot3}
\end{eqnarray}
Now we will rewrite the equation (\ref{eqrhot3}) above in term of
matrix elements of operators $\bar{\rho}(\tau)$ and $U(\phi;\tau,t)$
\begin{eqnarray}
\bar{\rho}(\tau)_{im}e^{\tau/\tau_{0}} & = &
\int dq(\phi_{0})\sum_{k,l}U(\phi_{0};\tau,0)_{ik}\rho(0)_{kl}U^{-1}(\phi_{0};\tau,0)_{lm}\nonumber\\
& + & \frac{1}{\tau_{0}}\int_{0}^{\tau}dt e^{t/\tau_{0}}\int
dq(\phi)
\sum_{k,l}U(\phi;\tau,t)_{ik}\bar{\rho}(t)_{kl}U^{-1}(\phi;\tau,t)_{lm}.\label{eqrhotm}
\end{eqnarray}
Define the conjunct of matrix $\{\mathbb{G}^{im}(\tau,t)\}$ with
elements give by
\begin{eqnarray}
G^{im}(\tau,t)_{lk} & = & \int dq(\phi)
U_{ik}(\phi;\tau,t)U_{lm}^{-1}(\phi;\tau,t),\label{eqGlk}
\end{eqnarray}
so
\begin{eqnarray}
\bar{\rho}(\tau)_{im}e^{\tau/\tau_{0}} & = &
\sum_{l}\Big[\sum_{k}G^{im}(\tau,0)_{lk}\rho(0)_{kl}\Big]\nonumber\\
& + & \frac{1}{\tau_{0}}\int_{0}^{\tau}dt e^{t/\tau_{0}}
\sum_{l}\Big[\sum_{k}G^{im}(\tau,t)_{lk}\bar{\rho}(t)_{kl}\Big]\nonumber\\
& = &
\sum_{l}[\mathbb{G}^{im}(\tau,0)\rho(0)]_{ll}\nonumber\\
& + & \frac{1}{\tau_{0}}\int_{0}^{\tau}dt e^{t/\tau_{0}}
\sum_{l}[\mathbb{G}^{im}(\tau,t)\bar{\rho}(t)]_{ll} .\label{eqrhotm}
\end{eqnarray}
Using the trace definition $TrA = \sum_{l}a_{ll}$ we get
\begin{eqnarray}
\bar{\rho}(\tau)_{im} & = &
e^{-\tau/\tau_{0}}Tr\big[\mathbb{G}^{im}(\tau,0)\rho(0)\big]\nonumber\\
& + & \frac{1}{\tau_{0}}\int_{0}^{\tau}dt
e^{-(\tau-t)/\tau_{0}}Tr\big[\mathbb{G}^{im}(\tau,t)\bar{\rho}(t)\Big].\label{eqrhot4}
\end{eqnarray}
This is the statistical average over the random variable $\phi(t)$.
In order to determine the dynamical evolution of the system one has
to determine $\mathbb{G}$. The problem is now simplified because we
have to deal with interval in which $\phi$ (or alternatively $g$) is
constant and the change in $\rho$ is perfectly regular. Thus,
knowing $\mathbb{G}$, we can find the average variation of the
system during the relaxation process \cite{bursh66}. To determine
$\mathbb{G}$ we will use the time evolution operator $U(t)$ given by
(\ref{Umtx}) with $g(t)$ defined in (\ref{gt})
\begin{eqnarray}
U(t) & = & \left(
          \begin{array}{cc}
            \cos\big(|g|t\big) & e^{i\phi}\sin\big(|g|t\big) \\
            -e^{-i\phi}\sin\big(|g|t\big) & \cos\big(|g|t\big) \\
          \end{array}
        \right).
\end{eqnarray}
The elements to be averaged in the calculations of $\mathbb{G}$ are
those containing the factors $e^{\pm i\phi}$. Since the phases are
equally probable most of the terms vanish after averaging. The
remaining (relevant for our purposes) non-vanishing elements of
$\mathbb{G}$ are
\begin{eqnarray}
G_{11}^{11} & = & G_{22}^{22} = \cos^{2}\big[(g_{0}(\tau
-t)\big]\nonumber\\
G_{22}^{11} & = & G_{11}^{22} = \sin^{2}\big[(g_{0}(\tau
-t)\big]\\
G_{21}^{12} & = & G_{12}^{21} = \cos^{2}\big[(g_{0}(\tau -t)\big].
\nonumber
\end{eqnarray}
We have thus
\begin{eqnarray}
G^{11}(\tau,t) & = & \left(
  \begin{array}{cc}
    \cos^{2}\big[(g_{0}(\tau
-t)\big] & 0 \\
    0 & \sin^{2}\big[(g_{0}(\tau
-t)\big] \\
  \end{array}
\right)\nonumber\\
\nonumber\\
G^{12}(\tau,t) & = & \left(
  \begin{array}{cc}
    0 & 0 \\
    \cos^{2}\big[(g_{0}(\tau
-t)\big] & 0 \\
  \end{array}
\right) = \big[G^{21}(\tau,t)\big]^{T}\label{eqG}
\\
\nonumber\\
 G^{22}(\tau,t) & = & \left(
  \begin{array}{cc}
    \sin^{2}\big[(g_{0}(\tau
-t)\big] & 0 \\
    0 & \cos^{2}\big[(g_{0}(\tau
-t)\big] \\
  \end{array}
\right),\nonumber
\end{eqnarray}
and using equations (\ref{eqrhot4}) and (\ref{eqG}) we obtain
\begin{eqnarray}
\bar{\rho}(\tau)_{11}e^{\tau/\tau_{0}} & = &
\rho(0)_{11}+\big[\rho(0)_{22}-\rho(0)_{11}\big]\sin^{2}\big(g_{0}\tau\big)\nonumber\\
& + & \frac{1}{\tau_{0}}\int_{0}^{\tau}dt
e^{t/\tau_{0}}\big\{\bar{\rho}(t)_{11}+\big[\bar{\rho}(t)_{22}
-\bar{\rho}(t)_{11}\big]\sin^{2}\big[g_{0}(\tau -
t)\big]\big\},\label{eqrhot5}\nonumber\\
\\
\bar{\rho}(\tau)_{22}e^{\tau/\tau_{0}} & = &
\rho(0)_{22}+\big[\rho(0)_{11}-\rho(0)_{22}\big]\sin^{2}\big(g_{0}\tau\big)\nonumber\\
& + & \frac{1}{\tau_{0}}\int_{0}^{\tau}dt
e^{t/\tau_{0}}\big\{\bar{\rho}(t)_{22}+\big[\bar{\rho}(t)_{11}
-\bar{\rho}(t)_{22}\big]\sin^{2}\big[g_{0}(\tau -
t)\big]\big\}.\nonumber
\end{eqnarray}

\section{Dynamics of the average number of a mode}
The average bosons number is given by
\begin{eqnarray}
n_{a}(t) = Tr\Big[\rho(t)a^{\dag}a\Big]
\end{eqnarray}
and can be evaluated using equations (\ref{eqrhot5}):
\begin{eqnarray}
\langle\overline{n_{a}(\tau)}\rangle e^{\tau/\tau_{0}} & = & \langle
n_{a}(0)\rangle\Big[1 - 2\sin^{2}\big(g_{0}\tau\big)\Big] + \langle
N \rangle \sin^{2}\big(g_{0}\tau\big)\nonumber\\
& + & \frac{1}{\tau_{0}}\int_{0}^{\tau}e^{t/\tau_{0}}\langle
\overline{n_{a}(t)}\rangle\Big\{1 -
2\sin^{2}\big[g_{0}(\tau-t)\big]\Big\}dt\label{eqn}\\
 & + &
\frac{\langle N\rangle}{\tau_{0}}\int_{0}^{\tau}e^{t/\tau_{0}}
\sin^{2}\big[g_{0}(\tau-t)\big]dt,\nonumber
\end{eqnarray}
where $\langle N \rangle = \langle n_{a}(0)\rangle + \langle
n_{b}(0)\rangle$ is the initial excitation number with $\langle
n_{a}(0)\rangle$ and $\langle n_{b}(0)\rangle$ being the average
excitations in each mode. This equation describes the relaxation of
the intensity of the mode $a$ and can be solved using the Laplace
transform. To solve (\ref{eqn}) we will define
\begin{eqnarray}
f(t) & = & \langle\overline{n_{a}(t)}\rangle
e^{t/\tau_{0}},\nonumber\\
g(t) & = & 1 - 2\sin^{2}\big(g_{0}t\big) =
\cos\big(2g_{0}t\big),\nonumber\\
h(t) & = & e^{t/\tau_{0}},\nonumber\\
j(t) & = & \sin^{2}\big(g_{0}t\big).\nonumber \label{eqfLP}
\end{eqnarray}
inserting the functions defined above into (\ref{eqn}) we obtain
\begin{eqnarray}
f(\tau) & = & \langle n_{a}(0)\rangle g(\tau) +
\frac{1}{\tau_{0}}\int_{0}^{\tau}f(t)g(\tau-t)dt\nonumber\\
& + & \langle N \rangle j(\tau) + \frac{\langle
N\rangle}{\tau_{0}}\int_{0}^{\tau}h(t)j(\tau-t)dt. \label{eqnLP}
\end{eqnarray}
Now applying the Laplace transform on both sides of (\ref{eqnLP}) we
obtain
\begin{eqnarray}
\hat{f}(s) & = & \langle n_{a}(0)\rangle\hat{g}(s) +
\frac{1}{\tau_{0}}\hat{f}(s)\hat{g}(s)\nonumber\\
& + & \langle N \rangle\hat{j}(s) + \frac{\langle
N\rangle}{\tau_{0}}\hat{h}(s)\hat{j}(s)\label{eqnLPT1}\\
& = & \frac{\langle n_{a}(0)\rangle\hat{g}(s)+\langle N \rangle
\hat{j}(s)+\frac{\langle N\rangle}{\tau_{0}}\hat{h}(s)\hat{j}(s)}
{\Big[1-\frac{1}{\tau_{0}}\hat{g}(s)\Big]}.\nonumber
\end{eqnarray}
Calculating the Laplace transform of the functions $\hat{g}(s)$,
$\hat{h}(s)$ and $\hat{j}(s)$ and substituting it into
(\ref{eqnLPT1}) we obtain
\begin{eqnarray}
\hat{f}(s) & = & \frac{\langle n_{a}(0)\rangle s}{\Big(s -
\frac{1}{2\tau_{0}}\Big)^{2}+\Omega^{2}} + \frac{\langle
N\rangle\big(2g_{0})^{2}}{2\Big(s -
\frac{1}{\tau_{0}}\Big)}\frac{1}{\Big[\Big(s -
\frac{1}{2\tau_{0}}\Big)^{2} + \Omega^{2} \Big]}, \label{eqnLPT2}
\end{eqnarray}
where we defined $\Omega = \sqrt{(2g_{0})^{2} -
\frac{1}{(2\tau_{0})^{2}}}$. The inverse Laplace transform of
Eq.(\ref{eqnLPT2}) yields the following expression the time
evolution of the relaxation as
\begin{eqnarray}
\overline{\langle n_{a}(t)\rangle} & = & \langle n_{a}(0)\rangle
e^{-t/2\tau_{0}}\Big[\cos\big(\Omega t\big) + \frac{\sin\big(\Omega
t\big)}{2\tau_{0}\Omega}\Big]\nonumber\\
& + & \frac{\langle N\rangle
e^{-t/2\tau_{0}}}{2}\Big\{\Big[e^{t/2\tau_{0}}- \cos\big(\Omega
t\big)\Big]-\frac{\sin\big(\Omega t\big)}{2\tau_{0}\Omega}\Big\}.
\label{eqnm}
\end{eqnarray}
In the limit when the average time $\tau_{0}$ between phase jumps is
large as compared with the oscillation period one obtains for
$g_{0}\tau_{0}\rightarrow \infty$ a pure oscillatory regime with
frequency $2g_{0}$. As the average time between frequency jumps
increases one obtains an envelop limiting the oscillation amplitudes
and the oscillation frequency is only slightly altered. However in
the limit where the average time between jumps decreases as compared
with the coupling $g_{0}$ the oscillation ceases at $g_{0}\tau_{0} =
0.25$ and an over damped limit sets in. Figures (\ref{fig2}) and
(\ref{fig3}) illustrate the weak coupling regime ($g_{0}\tau_{0}\gg
0.25$) and strong coupling regime ($g_{0}\tau_{0}
> 0.25$) respectively. Figure (\ref{fig4}) illustrates the ``freezing" of the
dynamics as $g_{0}\tau_{0}$ decreases beyond the limit
$g_{0}\tau_{0} = 0.25$. The fluctuating interact records information
of the system. The information transfer plays the role of unobserved
detection process \cite{zurek82}. In our case small values of
$g_{0}\tau_{0}$ would imply according to this reasoning that the
system is being ``measured" with increasing frequency, freezing out
as a Zeno like effect.

\section{Applications}
\subsection{The weak coupling regime: Exciton-polariton oscillations}
\begin{figure}[h]
\centering
\includegraphics[scale=0.35, angle=270]{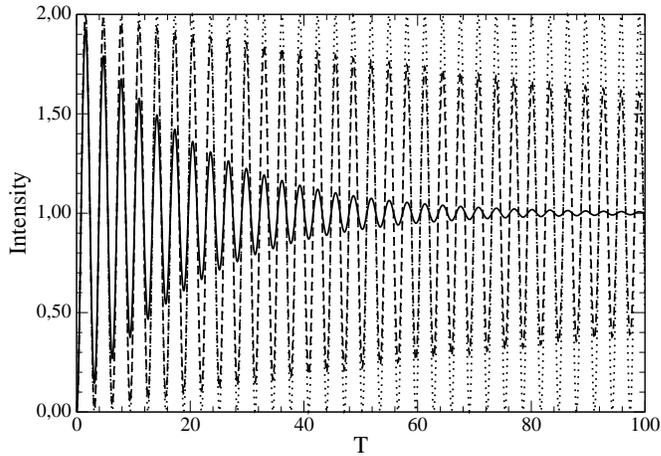}
\caption{Light intensity as a function of dimensionless time $T =
g_{0}t$ for the case that the excitons are initially in a number
state $N = 2$ for $g_{0}\tau_{0} \rightarrow\infty$ (dotted line),
$g_{0}\tau_{0} = 100$ (dashed line) and $g_{0}\tau_{0} = 10$ (solid
line). The intensity is in arbitrary units.} \label{fig2}
\end{figure}

\begin{figure}[h]
\centering
\includegraphics[scale=0.35, angle=270]{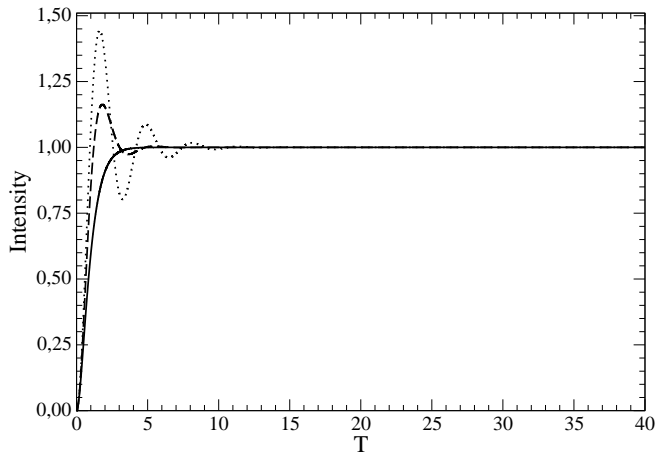} \caption{Light
intensity as a function of dimensionless time $T = g_{0}t$ for the
case that the excitons are initially in a number state $N_{b} = 2$
for $g_{0}\tau_{0} = 1.0$ (dotted line), $g_{0}\tau_{0} = 0.50$
(dashed line) and $g_{0}\tau_{0} = 0.25$ (solid line). The intensity
is in arbitrary units.}\label{fig3}
\end{figure}

Let us consider the weak coupling regime (WCR), where $g_{0}\tau_{0}
> 0.25$, and consequently $\Omega$ is a real number. Note that, as
$\tau_{0} \rightarrow \infty$, the expression for $\overline{\langle
n_{a}(t)\rangle}$ reduces to the usual result without fluctuations.
The effects of phase fluctuation in the intensity of the mode $a$ ,
e.g, for an initial number state $|\varphi(0)\rangle_{b} =
|N_{b}\rangle$ and the mode $a$ in vacuum state are given, according
to (\ref{eqnm}), by
\begin{eqnarray}
\overline{\langle n_{a}(t)\rangle} & = & \frac{N_{b}
e^{-t/2\tau_{0}}}{2}\Big\{\Big[e^{t/2\tau_{0}}- \cos\big(\Omega
t\big)\Big]-\frac{\sin\big(\Omega t\big)}{2\tau_{0}\Omega}\Big\}.
\label{eqnm1}
\end{eqnarray}
The result above shows that the intensity of the mode $a$ contains
two parts: (1) Rabi oscillations with frequency $\Omega$; (2) a
comparatively slow-varying part $e^{-t/2\tau_{0}}$. To see this more
clearly, we plot equation (\ref{eqnm1}) in Figs. (\ref{fig2}) and
(\ref{fig3}) for $N_{b} = 2$. It is clear that the damping of Rabi
oscillations is more pronounced when the mean time interval
$\tau_{0}$ between phase jumps become shorter and shorter until
$g_{0}\tau_{0} = 0.25$ (see Figures (\ref{fig2}) and (\ref{fig3})).
In other words the decoherence mechanism is faster for shorter-phase
jump intervals.

In this section we compare the results obtained in Ref. \cite{pau95}
which describe the exciton-polariton oscillations in the weak
coupling regime with a phenomenological coupling constant. In a real
cavity, the modes of exciton and photon are coupled to a continuum
of modes which leads to dissipation. The coupling can be scattering
of phonons in the case of exciton or cavity damping in the case of
photon. In both cases, the result is to dampen the mode of interest.
The result obtained in Ref. \cite{pau95} can be read of
Eq.(\ref{eqnm1}) where the damping coefficient corresponds to
$\tau_{0} = \gamma^{-1} = (\gamma_{p}+\gamma_{ex})^{-1}$, where
$\gamma_{ex}$ and $\gamma_{p}$ are exciton and photon damping from
the reservoirs. Using the values of $\gamma$ adopted in Ref.
\cite{pau95} we may estimate the order of magnitude of $\tau_{0}$ as
$10^{-10}s$.    The model has been shown to reproduce experimental
results \cite{yama93, weisb92}. In this situation equation
(\ref{eqnm1}) can be written as
\begin{eqnarray}
\overline{\langle n_{a}(t)\rangle} & = & \frac{N_{b}}{2}\Big[1-
e^{-\gamma t/2}\cos\big(2g_{0}t\big)\Big]. \label{eqnm3}
\end{eqnarray}
The results obtained here describe the decoherence process in the
system. However, the fluctuations introduce a finite width in the
transmission spectrum even in a lossless cavity. On the other hand,
in recent work, S. Schneider et al., \cite{schneider98} also
included fluctuation in intensity and phase in the exciting laser
pulse to explain effects of decoherence for single trapped ion. In
Schneider's model the intensity and phase fluctuations define a
stochastic Schr\"odinger equation in the Ito formalism
\cite{dyrt96}, or more appropriately a stochastic Liouville-von
Neumann equation. The results are in good qualitative agreement with
recent ion experiments \cite{meekhof96}.
\\
\\
\\
\\
\subsection{An alternative self-trapping mechanism}
\begin{figure}[h]
\centering
\includegraphics[scale=0.35, angle=270]{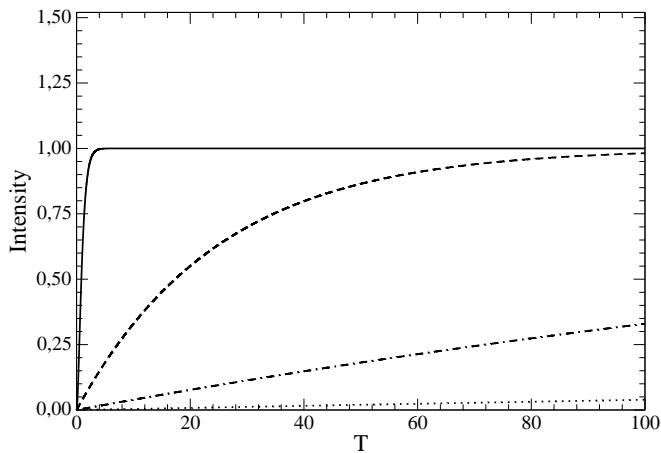}
\caption{Light intensity as a function of dimensionless time $T =
g_{0}t$ for the case that the excitons are initially in a number
state $N_{b} = 2$ for $g_{0}\tau_{0} = 0.25$ (solid line),
$g_{0}\tau_{0} = 0.01$ (dashed line), $g_{0}\tau_{0} = 0.001$
(dashed dotted line) and $g_{0}\tau_{0} = 0.0001$ (dotted line). The
intensity is in arbitrary units.} \label{fig4}
\end{figure}

Now we analyze the strong coupling regime (SCR), where
$g_{0}\tau_{0} < 0.25$. In this case $\Omega$ is purely imaginary
and Eq. (\ref{eqnm}) can be written as
\begin{eqnarray}
\overline{\langle n_{a}(t)\rangle} & = & \frac{N_{b}
e^{-t/2\tau_{0}}}{2}\Big\{\Big[e^{t/2\tau_{0}}- \cosh\big(|\Omega|
t\big)\Big]-\frac{\sinh\big(|\Omega|
t\big)}{2\tau_{0}|\Omega|}\Big\}. \label{eqnZeno}
\end{eqnarray}
The SCR may be investigated by looking at the intensity of the mode
$a$. As observed above when $g_{0}\tau_{0}$ becomes shorter and
shorter as compared to $0.25$, the fluctuation effects are larger,
and the fluctuations prevail over the oscillation between mode $a$
and mode $b$. In this case the SCR modifies the picture. The
inhibition of the transition of excitations between the modes is
induced by the fluctuations in the coupling. This can be interpreted
as an environment induced ``quantum Zeno-like effect (QZLE)"
\cite{rossi09, manis08, jgoliveira08, pascazio94, khodor08,
khodor09}. In the regime $g_{0}\tau_{0}\ll 0.25$ the interaction
between mode $a$ and mode $b$ is not able to absorb or release
energy and therefore stay put. The fluctuations in the interaction
strength between the mode $a$ and mode $b$ inhibits the excitation
of mode $a$ (in Fig. (\ref{fig4}), we exemplify this effect). When
$g_{0}\tau_{0} \rightarrow 0$, $\overline{\langle
n_{a}(t)\rangle}\rightarrow 0$, when $\langle n_{a}(0)\rangle = 0$;
the dynamics is frozen. In Ref.\cite{michael05} a self-trapping
mechanism of BEC in a laser potential has been reported. Two
explanations have been given. Firstly the one using a nonlinear
Gross-Pitayesty equation \cite{smerzi97, meier01} and the other a
schematic many body system \cite{salgueiro07}. In the present
contribution one might view modes $a$ and $b$ as the two sides of
the well and the self-trapping mechanism as the freezing out of the
dynamics due to uncontrollable fluctuations in the experiment.

\section{Conclusion}
We studied a system of two linearly coupled oscillators and the
effects of a phase fluctuating coupling. The model can be solved
analytically and displays the weak-strong coupling transition. We
show that this transition is a function of a dimensionless parameter
$g_{0}\tau_{0}$ and occurs at $g_{0}\tau_{0} = 0.25$. In the weak
coupling regime we provide for an analytical expression for the
damping parameter and compare with that of Ref. \cite{pau95}, in the
context of exciton-polariton oscillations. The strong coupling
regime leads to a ``freezing" of the dynamics and may qualitatively
provide for yet a third explanation for the self-trapping phenomenon
in BEC (the first two given in Refs. \cite{smerzi97, meier01} and
\cite{salgueiro07}).

\ack D.S.F and M.C.N acknowledge the financial support from Conselho
Nacional de Desenvolvimento Cientifico e Tecnol\'ogico - CNPq
(150232/2012-8), Brazil.

\end{document}